\documentclass[journal]{IEEEtran}
\usepackage{graphicx}
\usepackage{latexsym}
\usepackage{amsmath}
\usepackage{amsthm}
\usepackage{makecell,rotating,multirow,diagbox}
\usepackage{amsfonts}
\usepackage{subfigure}
\usepackage{dsfont}
\usepackage{epstopdf}
\usepackage{threeparttable}
\usepackage{multirow}

\usepackage{enumerate}

\usepackage{epsfig}

\usepackage{bm}

\usepackage{cite}

\usepackage{algorithmic}

\usepackage{textcomp}
\usepackage{xcolor}
\usepackage{algorithm}
\usepackage{algorithmic}
\usepackage{setspace}

\usepackage{amssymb}
\usepackage{amsmath}

\ifCLASSINFOpdf
\else
\fi
\begin{document}

%\title{Reconfigurable Intelligent Surface-Enhanced Channel Estimation Based on ELM Network with Insufficient Cyclic Prefix}
\title{Enhanced ELM Based Channel Estimation for RIS-Assisted OFDM systems with Insufficient \\ CP and Imperfect Hardware}
\author{{Chaojin~Qing,~\IEEEmembership{Member,~IEEE,}
       Li~Wang,
       Lei~Dong
       and~Jiafan~Wang}
%\author{Author~1,
%       Author~2,
%       Author~3
%       and~Author~4}
 \thanks{This work is supported in part by the Sichuan Science and Technology Program (Grant No. 2021JDRC0003), the Major Special Funds of Science and Technology of Sichuan Science and Technology Plan Project (Grant No. 19ZDZX0016 /2019YFG0395), the Demonstration Project of Chengdu Major Science and Technology Application (Grant No. 2020-YF09-00048-SN), the Key Scientific Research Fund of Xihua University (Grant No. Z1120941), and the Special Funds of Industry Development of Sichuan Province (Grant No. zyf-2018-056).}

\thanks{C. Qing, L. Wang and L. Dong are with the School of Electrical Engineering and Electronic Information, Xihua University, Chengdu, 610039, China (E-mail: qingchj@mail.xhu.edu.cn). }
\thanks{J. Wang is with the Synopsys Inc., 2025 NE Cornelius Pass Rd, Hillsboro, OR 97124, USA (E-mail: jifanw@gmail.com).}
}

\markboth{IEEE XXXXXXXX XXXXXXX,~Vol.~XX, No.~XX, XXX~2021}%
 {Shell \MakeLowercase{\textit{et al.}}: Bare Demo of IEEEtran.cls for IEEE Journals}

\maketitle

\begin{abstract}
Reconfigurable intelligent surface (RIS)-assisted orthogonal frequency division multiplexing (OFDM) systems have aroused extensive research interests due to the controllable communication environment and the performance of combating multi-path interference. However, as the premise of RIS-assisted OFDM systems, the accuracy of channel estimation is severely degraded by the increased possibility of insufficient cyclic prefix (CP) produced by extra cascaded channels of RIS and the nonlinear distortion lead by imperfect hardware. To address these issues, an enhanced extreme learning machine (ELM)-based channel estimation (eELM-CE) is proposed in this letter to facilitate accurate channel estimation. Based on the model-driven mode, least square (LS) estimation is employed to highlight the initial linear features for channel estimation. Then, according to the obtained initial features, an enhanced ELM network is constructed to refine the channel estimation. In particular, we start from the perspective of guiding it to recognize the feature, and normalize the data after the network activation function to enhance the ability of identifying non-linear factors. Experiment results show that, compared with existing methods, the proposed method achieves a much lower normalized mean square error (NMSE) given insufficient CP and imperfect hardware. In addition, the simulation results indicate that the proposed method possesses robustness against the parameter variations.
\end{abstract}


\section{Introduction}
\IEEEPARstart{R}{econfigurable} intelligent surface (RIS) has been recognized as a potential technology for future sixth generation (6G) mobile communications \cite{c2}, due to its promising performances in spectrum efficiency (SE), energy efficiency (EE), etc \cite{c25}. In particular, the wireless propagation environment is manipulated by changing the phase shift \cite{c24}, and this characteristic is employed for ``smart cities'' in massive machine type communication (mMTC) \cite{c3}. To support the performance enhancement of wireless communication systems, e.g., the SE and EE, accurate channel estimation is the premise of RIS systems \cite{c13}, \cite{c1}. However, the channel estimation in RIS-assisted wireless systems is quite challenging due to the prohibitively high pilot overhead \cite{c2}, complicated cascaded channels \cite{c1} and passive surface elements \cite{c22}, etc. Especially, the premise of RIS, e.g., accurate channel estimation, is suffered from multi-path interference. Given that the orthogonal frequency division multiplexing (OFDM) is good at resisting multi-path interference \cite{c15}, the RIS-assisted OFDM systems are taken into account in this letter.

%The orthogonal frequency division multiplexing (OFDM) is good at resisting multi-path interference \cite{c15}, thus the RIS-assisted OFDM systems make sense \cite{c13}, \cite{c1}.
%In various communication standards, e.g., the fourth generation (4G) and fifth generation (5G) mobile communications, the orthogonal frequency division multiplexing (OFDM) has been adopted \cite{c15} to combat multi-path interference. Thus, RIS-assisted OFDM systems have attracted extensive attention \cite{c13}, \cite{c1}.

For RIS-assisted OFDM systems, existing researches, e.g., \cite{c23,c24,c25,c26}, are studied under two assumptions. First, the cyclic prefix (CP) length in RIS-assisted OFDM systems is longer than the maximum delay spread. Second, perfect hardware is applied in the systems. However, these two assumptions are not common for practical application in RIS-assisted OFDM systems. For the first assumption, namely, the sufficient CP, challenges come from the time-varying channels and RIS-introduced extra paths. Specifically, adopting a sufficient CP to deal with time-varying scenarios may consume massive spectrum resources \cite{c4}. Given extra paths introduced by RIS, the possibility will increase that the previously indiscernible path becomes a resolvable path, and this, in turn, increases the possibility of insufficient CP \cite{c18}. Therefore, to save valuable spectrum resources, the development of receivers with insufficient CP is highly desired.
%For example, the extra wireless paths and the increased probability of maximum delay of paths produced by RIS usually cause insufficient CP with high probability \cite{c18}, and the CP length is usually fixed while the propagation environment is time-varying. Furthermore, a longer CP means a lower SE because the CP does not contain additional information \cite{c4}.
In addition to the insufficient CP, imperfect hardware is usually observed in most practical RIS-assisted OFDM systems \cite{c19}. For example, high power amplifier (HPA), digital to analog converter (DAC), etc., these hardware inevitably causes nonlinear distortion in the real application scenarios. For channel estimation, the nonlinear distortion destroys the orthogonality of the training sequence and seriously degrades the estimation accuracy \cite{c17}. Therefore, previous works, e.g., \cite{c24,c25} and \cite{ c1}, mainly based on the previous two assumptions, could not work well in practical applications given the scenarios of insufficient CP and nonlinear distortion.

%In many previous works, e.g., \cite{c24,c25} and \cite{ c1}, the insufficient CP and nonlinear distortion have not been considered, causing them cannot work very well in these application scenarios.

%In summary, insufficient CP and nonlinear distortion have not been considered in many previous works, e.g., \cite{c24,c25}, causing them cannot work very well.
%
%this assumption is not practical. For example, the increased wireless paths produced by RIS usually cause insufficient CP with high probability \cite{c18}, and the CP length is usually fixed while the propagation environment is time-varying. Especially, a longer CP means a lower SE because the CP does not contain additional information \cite{c4}.
%%Although a scenario without CP appears, it requires over-sampling, thereby increasing the receiver's cost and complexity  \cite{c5}.
%In contrast, the scenarios of insufficient CP are more common for the practical application in RIS-assisted OFDM systems, which make a tradeoff between bandwidth efficiency and complexity, and possess better flexibility.
%
%In addition to the insufficient CP, imperfect hardware is usually observed in most practical RIS-assisted OFDM systems \cite{c19}, which causes serious nonlinear distortion and limits the system performance \cite{c16}. For channel estimation, the nonlinear distortion destroys the orthogonality of the training sequence \cite{c17}, seriously degrading the estimation accuracy.

To improve the channel estimation of RIS-assisted OFDM systems affected by insufficient CP and imperfect hardware, we introduce an enhanced extreme learning machine (ELM) network. Unlike the deep learning-based methods that require complex parameter tuning and long training time \cite{c21}, the proposed ELM has the advantages of fast learning speed and performance robustness. We enhance the ELM network by using hidden-layer standardization and propose an enhanced ELM-based channel estimation (eELM-CE) method for RIS-assisted OFDM systems. Different from the common standardization method, our standardization is added before the activation function of hidden layer to ease feature acquirement for ELM's hidden layer output. In the proposed method, the imperfect hardware and the insufficient CP are modeled as a nonlinear problem and solved by exploiting the learning ability of the enhanced ELM. Generally, the proposed method can be viewed as the model-driven mode \cite{c0} integrated with the least square (LS) estimation and an enhanced ELM network, which extracts initial features and refines the following channel estimation, respectively. In contrast, \cite{c24,c25} are mainly data-driven, without the cooperation of initial features. Simulation results show that the performance of channel estimation is significantly improved by using our proposed eELM-CE method. To the best of our knowledge, this is the first work that nonlinear distortion with insufficient CP is taken into account in channel estimation for RIS-assisted OFDM systems.

The remainder of this letter is structured as follows: In Section II, the system model of RIS-assisted OFDM systems with insufficient CP and imperfect hardware is presented. The proposed method, i.e., eELM-CE is presented in Section III, and numerical results are given in Section IV. Finally, Section V concludes our work.
% Note that the IEEE does not put floats in the very first column
% - or typically anywhere on the first page for that matter. Also,
% in-text middle ("here") positioning is typically not used, but it
% is allowed and encouraged for Computer Society conferences (but
% not Computer Society journals). Most IEEE journals/conferences use
% top floats exclusively.
% Note that, LaTeX2e, unlike IEEE journals/conferences, places
% footnotes above bottom floats. This can be corrected via the
% \fnbelowfloat command of the stfloats package.

\textit{Notations}: Boldface upper case and lower case letters denote matrix and vector respectively. ${\left(\cdot \right)^T}$ and $\mathbf{(\cdot)^\dag}$ denote the transpose and matrix pseudo-inverse, respectively. $ \odot $ is the Hadamard product. ${\left\|  \cdot  \right\|_2}$ is the Euclidean norm. ${\mathbf{F}_N}$ is the normalized $N \times N$ Fourier transform matrix. ${{\bf{0}}_N}$ is an $N \times 1$ zero vector. $\mathds{E}(\cdot)$ is the mathematical expectation. $\mathds{D}(\cdot)$ represents the variance.

\section{System Model}
%The conclusion goes here.
We consider a RIS-assisted OFDM system, which comprises one single-antenna transmitter, one single-antenna receiver and one multi-element RIS. This system employs $N$ subcarriers, and assumes the CP length ${L_{{\rm{\mathrm{CP}}}}}$ is shorter than the maximum delay spread $L$, i.e., ${L_{{\rm{\mathrm{CP}}}}} < L$.
The time-domain signal received at the receiver is given by
\begin{equation}\label{EQ1}
{\bf{y}} = {{\bf{\widetilde{X}}}_c}{\bf{h}} + {\bf{n}},
\end{equation}
where $ {{\bf{\mathbf{\widetilde{X}}}}_\mathrm{c}} \in \mathbb{C}^{{\left( {N + {L_{\mathrm{CP}}} + L - 1} \right) \times L}}$ is a cyclic matrix formed by the transmitted data, $\mathbf{h}={\left[ {{h_1},{h_2}, \cdots ,{h_L}} \right]^T}$ stands for the composite channel impulse response (CIR) between the receiver and transmitter, and ${\bf{n}} \in \mathbb{C}^{\left( {N + {L_{\mathrm{CP}}} + L - 1} \right) \times 1}$ denotes the circularly symmetric complex Gaussian (CSCG) distribution with zero mean and variance ${\sigma ^2}$. In (\ref{EQ1}), the first column of ${{\bf{\mathbf{\widetilde{X}}}}_\mathrm{c}}$ is given as
\begin{equation}\label{EQ1_ADD1}
\widetilde{\mathbf{x}}= [ {{\widetilde{x}_{{- {L_{CP}} + 1}}}, \cdots ,{\widetilde{x}_{0}}}, {{\widetilde{x}_1}, \cdots ,{\widetilde{x}_N},}{{\bf{0}}_{L - 1}}]^T ,
\end{equation}
where ${\widetilde{x}_i}$, $i={- {L_{\mathrm{CP}}} + 1},\cdots ,{{N}}$ denotes the distorted data. That is, the transmitted data ${{x}_i}$ experiences the nonlinear distortion caused by imperfect hardware to form ${\widetilde{x}_i}$, which is expressed as
\begin{equation}\label{EQ1_ADD2}
{\widetilde{x}_i} = f_{\mathrm{dis}} \left ( {{x}_i} \right ),
\end{equation}
where we use the function $f_{\mathrm{dis}}(\cdot)$ to describe the nonlinear distortion caused by imperfect hardware\cite{c21}.
%\textcolor[rgb]{1.00,0.00,0.00}{low resolution ADC, ***HPA, etc [***]}.

%
% experiencing nonlinear distortion caused by imperfect hardware. The first column of ${{\bf{\mathbf{\widetilde{x}}}}_\mathrm{c}}$ is given as $\widetilde{\mathbf{x}}= [ {{\widetilde{x}_{{N - {L_{CP}} + 1}}}, \cdots ,{\widetilde{x}_{N}},}{{\widetilde{x}_1}, \cdots ,{\widetilde{x}_N},}{{\bf{0}}_{L - 1}}]^T $. $\mathbf{h}={\left[ {{h_1},{h_2}, \cdots ,{h_L}} \right]^T}$ stands for the composite channel impulse response (CIR) between the receiver and transmitter. ${\bf{n}} \in \mathbb{C}^{\left( {N + {L_{\mathrm{CP}}} + L - 1} \right) \times 1}$ denotes the circularly symmetric complex Gaussian distribution with zero mean and variance ${\sigma ^2}$.

The composite CIR between the receiver and transmitter is represented by
\begin{equation}\label{EQ2}
%{\bf{h}}\;{\rm{ = }}\;{{\bf{h}}_{\mathrm{\mathrm{URB}}}}\bm{\phi}  + {{\bf{h}}_{\mathrm{UB}}},
{\bf{h}}\;{\rm{ = }}\;{{\bf{h}}_{\mathrm{TR}}} + {{\bf{H}}_{\mathrm{\mathrm{TRR}}}}\bm{\phi},
\end{equation}
where ${{\bf{h}}_{\mathrm{\mathrm{TR}}}} \in \mathbb{C}^{L \times 1}$ is the CIR of the transmitter-receiver direct link, ${{\bf{H}}_{\mathrm{TRR}}} \in \mathbb{C}^{L \times M}$ is the equivalent cascaded CIR of the reflecting link, ${{\bf{H}}_{\mathrm{\mathrm{TRR}}}} = \left[ {{{\bf{h}}_{\mathrm{\mathrm{TRR}},1}},{{\bf{h}}_{\mathrm{\mathrm{TRR}},2}}, \cdots ,{{\bf{h}}_{\mathrm{\mathrm{TRR}},M}}} \right]$ is stacked by ${{\bf{h}}_{\mathrm{\mathrm{TRR}},m}}$ with $m = 1, \cdots ,M$, and $\bm{\phi}  \buildrel \Delta \over = {\left[ {{\phi _1},{\phi _2}, \cdots ,{\phi _M}} \right]^T}$ means the phase-shift vector defined by ${\phi _m} = {\beta _m}{e^{j{\varphi _m}}}$. Similar to \cite{c1}, in order to maximize the reflection power of the RIS and simplify the hardware design, we fix ${\beta _m} = 1$.

 The equivalent cascaded CIR of the reflecting link is given by
\begin{equation}\label{EQ3}
{{\bf{h}}_{\mathrm{TRR},m}} = {{\bf{h}}_{\mathrm{TR},m}} \odot {{\bf{h}}_{\mathrm{RR},m}},
\end{equation}
where ${{\bf{h}}_{\mathrm{TR},m}} \in \mathbb{C}^{L \times 1}$ and ${{\bf{h}}_{\mathrm{RR},m}} \in \mathbb{C}^{L \times 1}$ are the aggregated CIRs of the transmitter-RIS link and RIS-receiver link associated with the $m$-th sub-surface, respectively.

%is the aggregated CIR of the transmitter-RIS link associated with the $m$-th sub-surface. Correspondingly ${{\bf{h}}_{\mathrm{RR},m}} \in \mathbb{C}^{L \times 1}$ is the aggregated CIR of the RIS-receiver link associated with $m$-th sub-surface.

According to (\ref{EQ1}), (\ref{EQ2}) and (\ref{EQ3}), the equivalent baseband received signal in time domain is rewritten by
\begin{equation}\label{EQ4}
%{\bf{y}} = {{\bf{\widetilde{x}}}_c}\left( {\sum\limits_{m = 1}^M {{{\bf{h}}_{{\rm{\mathrm{TR}}},m}}{\phi _m} \odot {{\bf{h}}_{{\rm{\mathrm{RR}}},m}} + {{\bf{h}}_{{\rm{\mathrm{TR}}}}}} } \right) + {\bf{n}}.
{\bf{y}} = {{\bf{\widetilde{X}}}_c}\left({{\bf{h}}_{{\rm{\mathrm{TR}}}}}+ {\sum\limits_{m = 1}^M ({{{\bf{h}}_{{\rm{\mathrm{TR}}},m}}{\phi _m} \odot {{\bf{h}}_{{\rm{\mathrm{RR}}},m}}  } } )\right) + {\bf{n}}.
\end{equation}

With the received signal ${\bf{y}}$, we use the LS estimation to highlight the initial estimation features for easing network learning. Then, an eELM-CE method is proposed in Section III to improve the estimation accuracy for the case of insufficient CP with imperfect hardware.

%${\tilde x_i},i = 1,2, \cdots ,N$

\section{Enhanced ELM-based Channel Estimation}
\subsection{Estimation Preprocessing}
With the received signal ${\bf{y}}$, we first remove its CP, and then employ LS estimation to obtain the initial features of channel estimation in the frequency domain. According to the transmission protocol proposed in \cite{c11}, two OFDM blocks are transmitted in each time slot. From \cite{c11}, it is assumed that the pilot tone, denoted as ${\bf{c}}$, and the transmitted data consist in the first and second OFDM block, respectively. Hence, we transmit $M+1$ slots due to the purpose to separate the direct link and reflecting link \cite{c1}.

For LS estimation, the received pilot signal, denoted by ${\mathbf{Y}_\mathrm{P}} = \left[ {{\mathbf{y}_{\mathrm{P},1}},{\mathbf{y}_{\mathrm{P},2}}, \cdots ,{\mathbf{y}_{\mathrm{P},M + 1}}} \right]$, is first picked out from the received signal $\mathbf{Y} = \left[ {{\mathbf{y}_1},{\mathbf{y}_2}, \cdots ,{\mathbf{y}_{M + 1}}} \right]$. Then, the received pilot signal ${{\bf{Y}}_\mathrm{P}}$ and transmitted pilot tone $\mathbf{C} = \left[ {{\mathbf{c}_1},{\mathbf{c}_2}, \cdots ,{\mathbf{c}_{M + 1}}} \right]$ are respectively transformed into frequency domain by using Discrete Fourier Transform (DFT), which are expressed as

\begin{equation}\label{EQ5}
\left\{ \begin{array}{l}
{\!\mathbf{\overline{Y}}_\mathrm{p}} = {\mathbf{F}_N}{\mathbf{Y}_\mathrm{p}}\\
\mathbf{\overline{C}} \;\:= {\mathbf{F}_N}\mathbf{C}
\end{array} \right.,
\end{equation}
where ${{\bf{\overline{Y}}}_\mathrm{p}}$ and ${\bf{\overline{C}}}$ represent the received pilot signal and transmitted pilot in the frequency domain, respectively.
With ${{\bf{\overline{Y}}}_\mathrm{p}}$ and ${\bf{\overline{C}}}$, we employ LS estimation to obtain the initial channel frequency response (CFR) of the composite channel, which is given by
%the composite channel frequency response (CFR) obtained by LS pre-estimation is given by

\begin{equation}\label{EQ6}
{{{\widehat{H}}}_{{\mathrm{LS}}}{(m,n)}} = \frac{{{{{\overline{Y}}}_{\mathrm{p}}{(m,n)}}}}{{{{\overline{C}}}\,\,{(m,n)}}},m = 1, \cdots ,N;n = 1, \cdots ,M + 1
\end{equation}
 where $m$ represents the $m$-th row, $n$ represents the $n$-th column and ${{\bf{\widehat{H}}}_{\mathrm{LS}}}$ is the composite CFR between the receiver and transmitter. Using the transmission protocol proposed in \cite{c1}, the CFRs of direct link and reflecting link are separated by

\begin{equation}\label{EQ7}
\left[ {{{\bf{\widehat{h}}}_{\mathrm{TR}}}\;{{\bf{\widehat{H}}}_{\mathrm{TRR}}}} \right] = {{\bf{\widehat{H}}}_{\mathrm{LS}}}{\bm{\Theta} ^{ - 1}},
\end{equation}
where ${{\bf{\widehat{h}}}_{\mathrm{TR}}} \in \mathbb{C}{^{N \times 1}}$ denotes the direct-link CFR, ${{\bf{\widehat{H}}}_{\mathrm{TRR}}} \in \mathbb{C}^{N \times M}$ is the reflecting-link CFR composed of $M$ sub-surfaces reflection paths, and ${\bm{\Theta}}$ represents the RIS reflection matrix which is a full-rank matrix \cite{c1}. From \cite{c1}, ${\bm{\Theta}}$ is described as
% and the reflecting link is composed of M sub-surfaces reflection paths, ${\bm{\Theta}}$ denotes the RIS reflection matrix, which is described as
\begin{equation}\label{EQ8}
{\bf{\Theta }}{\kern 1pt} {\rm{ = }}\left[ \begin{array}{l}
\,1\quad \quad \,\,1\quad \quad  \,\,\cdots \quad \quad \;\,\,\:1\\
{\bm{\phi} ^{\left( 1 \right)}}\quad {\bm{\phi} ^{\left( 2 \right)}}\quad \; \cdots \quad \quad {\bm{\phi} ^{\left( {M + 1} \right)}}\quad
\end{array} \right],
\end{equation}
where ${\bm{\phi} ^{\left( i \right)}}$ is the RIS's phase shift of the $i$-th pilot symbol.

Based on the initial CFRs, i.e., ${{\bf{\widehat{h}}}_{\mathrm{TR}}}$ and ${{\bf{\widehat{H}}}_{\mathrm{TRR}}}$, an eELM-CE is proposed in the following subsection (i.e., Section III-B) to refine the channel estimation.
\begin{algorithm}[t]
\begin{spacing}{0.8}
%\end{spacing}
\caption{eELM-CE method}
\begin{spacing}{1.2}
\end{spacing}
\hspace*{0.02in}{\bf Input:}
Separated CFRs $\mathbf{\widehat{H}} \buildrel \Delta \over = [ {{\mathbf{\widehat{h}}_{\mathrm{TR}}},{\mathbf{\widehat{h}}_{\mathrm{TRR},1}},}$
$ {\cdots, {\mathbf{\widehat{h}}_{\mathrm{TRR},M}}} ]$.
%\hspace*{0.4in} label $\mathbf{H} \buildrel \Delta \over = \left[ {{\mathbf{H}_{\mathrm{TR}}},{\mathbf{H}_{\mathrm{TRR},1}}, \cdots , {\mathbf{H}_{\mathrm{TRR},M}}} \right]$.

%j = 1, \cdots ,M + 1
%
%the direct link ${{\bf{\widehat{H}}}_{\mathrm{TR}}}$ and reflecting link ${{\bf{\widehat{H}}}_{\mathrm{TRR}}}$, \\
%\hspace*{0.45in}denoted as

%$\mathbf{\widehat{H}} \buildrel \Delta \over= \left[ {{\mathbf{\widehat{H}}^{\left( 1 \right)}}, \cdots ,{\mathbf{\widehat{H}}^{\left( {{N_t}} \right)}}} \right],t = 1, \cdots ,{N_t}$
%\hspace*{0.42in} in which
%${{ \mathbf{{{\widehat{H}}}}}^{\left( t \right)}} \buildrel \Delta \over = [ {{\mathbf{{{\widehat{H}}}}^{\left( t \right)}_{\mathrm{TR}}}},{\mathbf{{{\widehat{H}}}}^{\left( t \right)}_{\mathrm{TRR},1}}, \cdots ,{{\mathbf{{{\widehat{H}}}}^{\left( t \right)}_{\mathrm{TRR},M}}} ]$.\\
\hspace*{0.02in}{\bf Output:}
Refined CFRs $\mathbf{\widetilde{H}} \buildrel \Delta \over =\left[ {{\mathbf{\widetilde{h}}_{\mathrm{TR}}},{\mathbf{\widetilde{h}}_{\mathrm{TRR},1}}, \cdots ,{\mathbf{\widetilde{h}}_{\mathrm{TRR},M}}} \right]$.

%$\mathbf{\widetilde{H}} \buildrel \Delta \over= \left[ {{\mathbf{{\widetilde{H}}}^{\left( 1 \right)}}, \cdots ,{\mathbf{{\widetilde{H}}}^{\left( {{N_t}} \right)}}} \right],t = 1,\cdots , $\\
%\hspace*{0.51in} $ {N_d}$ in which
%${{ \mathbf{{\widetilde{H}}}}^{\left( t \right)}} \buildrel \Delta \over = [ {{\mathbf{{\widetilde{H}}}^{\left( t \right)}_{\mathrm{TR}}}},{\mathbf{{\widetilde{H}}}^{\left( t \right)}_{\mathrm{TRR},1}}, \cdots ,{{\mathbf{{\widetilde{H}}}^{\left( t \right)}_{\mathrm{TRR},M}}} ]$.
\begin{spacing}{1.2}
\end{spacing}
\hspace*{0.03in}{\bf Offline Training:}
\begin{spacing}{-0.9}
\end{spacing}
\begin{algorithmic}[1]
%\STATE { Generate the label $\mathbf{H}$.}

%\hspace*{0.7in}{\bf Offline Training phase}
%{\noindent\textbf{Offline Training phase}\\}
\begin{spacing}{2.0}
\end{spacing}
\STATE { Initialize the weight $\mathbf{W}$ and bias $\mathbf{b}$.}
%\par \setlength\parindent{1em}
\par \setlength\parindent{1em}\FOR{$j=1, ..., M+1$}
\par \setlength\parindent{1em}\FOR{$t=1,...,N_\mathrm{d}$}

%\par \setlength\parindent{1em}\FOR{$t=1,...,N_\mathrm{t}$}
%\par \setlength\parindent{1em}\FOR{$j=1, ..., M+1$}
%\par \setlength\parindent{1em}\FOR{$t=1,...,N_\mathrm{t}$}
 %\STATE { Generate the labels ${{ \mathbf{{{H}}}^{\left( t \right)}}} \buildrel \Delta \over = [ {{\mathbf{{H}}^{\left( t \right)}_{\mathrm{TR}}}},{\mathbf{{H}}^{\left( t \right)}_{\mathrm{TRR},1}}, \cdots ,$
%${{\mathbf{{H}}^{\left( t \right)}_{\mathrm{TRR},M}}} ]$.}\\
\STATE {Normalize the samples $\mathbf{\widehat{h}}^{\left( t \right)}_j$ using Eq.~(\ref{EQ9})} to form $\mathbf{\widehat{h}}^{\left( t \right)}_{\mathrm{st},j}$.\\
%\ENDFOR\\
% to form the training set ${{ \mathbf{{\widetilde{H}}}}^{\left( t \right)}_{\mathrm{st}}} \buildrel \Delta \over = [ {{\mathbf{{\widetilde{H}}}^{\left( t \right)}_{\mathrm{TR}}}},{\mathbf{{\widetilde{H}}}^{\left( t \right)}_{\mathrm{TRR},1}}, $
% $ \cdots ,{{\mathbf{{\widetilde{H}}}^{\left( t \right)}_{\mathrm{TRR},M}}} ]$\\
%\STATE { Use the CFRs preprocessed and labels to form a training
% set $\left\{ {\left( {{{\bf{\widehat{H}}}_j},{{\bf{H}}_j}} \right)\left| {j = 1,2, \cdots ,M + 1} \right.} \right\}$.}

% \par \setlength\parindent{1em}\FOR{$j=1, ..., M+1$}

\STATE{ Calculate the output ${\mathbf{o}^{\left( t \right)}_j}$ of the hidden layer after
 ${\mathbf{\widehat{h}}^{\left( t \right)}_{\mathrm{st},j}}$
through the activation function $\sigma \left( \cdot \right)$ using
 Eq.~(\ref{EQ10}) and Eq.~(\ref{EQ11}) according to $\mathbf{W}$ and $\mathbf{b}$.}

\ENDFOR\\
%\STATE { Generate the label $\mathbf{H}_j$.}
\STATE Collect the output ${\bm{\mathrm{O}} _j} = \left[ {\bm{\mathrm{o}} _j^{\left( 1 \right)}, \cdots ,\bm{\mathrm{o}} _j^{\left( {{N_d}} \right)}} \right]$.
\STATE {Use the known labels ${\mathbf{H}_j}$ and ${\mathbf{O}_j}$ to calculate the
 hidden layer output weight ${\bm{\beta} _j}$ using Eq.~(\ref{EQ12}).}
\ENDFOR\\

\begin{spacing}{1.3}
\end{spacing}
\hspace*{-0.33in}{\bf Online Deployment:}
%\end{spacing}
\begin{spacing}{0.4}
\end{spacing}
 %{\STATE\noindent\textbf{Online Operation phase}}
%\par \setlength\parindent{1em}\FOR{$t=1,...,N_\mathrm{d}$}
%
%
%\ENDFOR
%\STATE Collect the sample ${\mathbf{\overline{H}}_{\mathrm{st}}} = \left[ \begin{array}{l}
%\:\mathbf{\overline{H}}_{\mathrm{st}}^{\left( 1 \right)}\\
%\;\,\, \:\vdots \\
%\mathbf{\overline{H}}_{\mathrm{st}}^{\left( {{N_d}} \right)}
%\end{array} \right]$.

%Form the online sample $\mathbf{\dot{H}} \buildrel \Delta \over = \left[ {{\mathbf{\dot{H}}_1}, \cdots ,{\mathbf{\dot{H}}_{M + 1}}} \right],j = 1, \cdots ,M + 1$

\STATE { Perform LS estimation to obtain ${{\bf{{\overline{h}}}}_{\mathrm{LS}}}$ using
 Eq. (\ref{EQ6}).}
\STATE{Separate the CFRs of the direct link ${{{\bf{{\overline{h}}}}_{\mathrm{TR}}}}$ and reflecting
 link ${{{\bf{{\overline{H}}}}_{\mathrm{TRR}}}}$ using Eq. (\ref{EQ7}).}
 \STATE {Use the CFRs preprocessed to form ${{ \mathbf{{{\overline{H}}}}}} \buildrel \Delta \over = [ {{\mathbf{{\overline{h}}}_{\mathrm{TR}}}},$
 ${\mathbf{{\overline{h}}}_{\mathrm{TRR},1}}, \cdots ,$
${{\mathbf{{\overline{h}}}_{\mathrm{TRR},M}}} ]$.}
\FOR{$j=1, ..., M+1$}
\STATE { Normalize ${{ \mathbf{{{{\overline{h}}}}}}_j}$ using Eq.~(\ref{EQ9})} to form ${{ \mathbf{{{\overline{h}}}}}_{\mathrm{st},j}} $.
%\STATE Pick out the $j$ column of ${\mathbf{\overline{H}}_{\mathrm{st}}}$ and reshape it into a matrix, named as ${{ \mathbf{{\dot{H}}}}_{j}}$.
\STATE { Feed ${{ \mathbf{{{\overline{h}}}}}_{\mathrm{st},j}} $ into the trained ELM-based network using
 Eq.~(\ref{EQ13}) to obtain the refined CFR ${{ \mathbf{{\widetilde{h}}}}_j}$. }

\ENDFOR
%\STATE Collect the output $\mathbf{\widetilde{H}} = \left[ \begin{array}{l}
%{\,\,\:\mathbf{\widetilde{H}}_1}\\
%\; \:\:\:\:\vdots \\
%{\mathbf{\widetilde{H}}_{M + 1}}
%\end{array} \right]$.
%\FOR{$t=1, ..., N_\mathrm{d}$}
%\STATE Pick out the $t$ column of $\mathbf{\widetilde{H}}$ and reshape it into the matrix like $\left[ {{\mathbf{\widetilde{H}}_{\mathrm{TR}}},{\mathbf{\widetilde{H}}_{\mathrm{TRR},1}}, \cdots ,{\mathbf{\widetilde{H}}_{\mathrm{TRR},M}}} \right]$.
%\ENDFOR
\end{algorithmic}
\end{spacing}
\end{algorithm}

\subsection{Refining Estimation}
The details of eELM-CE are presented in \textbf{Algorithm~1}. The processes of offline training and online deployment are respectively given as follows.

%$1)$ \textit{Offline Training}:
\subsubsection{Offline Training}
In the training phase of ELM-based network, the main task is to learn the output weight. For ELM-based network, we need to train $M+1$ ELM networks, and each ELM network training uses ${N_d}$ samples. With extensive experiments, it is found that $N_d = 10000$ training samples show a good enough learning effect while no obvious improvement is noticed given more training samples. $\mathbf{\widehat{h}}_j^{\left( t \right)}$ represents the CFRs of the $j$-th ELM network, $j=1, ..., M+1$, and the superscript $t$, $t=1,...,N_d$, is the number of training sessions. We send the samples which are obtained using Eq.~(9) into the network to calculate the network output and then obtain the output weight of hidden layer.
%${N_d}$ is the number of training set, the superscript $t$ represents the number of training sessions, and the subscript $j$ indicates the $j$-th network, there are a total of $M+1$ networks for separated $M+1$ paths, $\mathbf{\widehat{H}}_j $ in which $j=1,\cdots, M + 1$ represents the CFRs of the direct link, the first sub-surface of reflecting link, ..., the $M$-th sub-surface of reflecting link.
In the algorithm, we obtain the samples and labels through Eq.~(\ref{EQ6}) and Eq.~(\ref{EQ7}), and then normalize the samples, which is expressed as
%separated CFRs of the direct link and reflecting link
\begin{equation}\label{EQ9}
\mathbf{\widehat{h}}^{\left( t \right)}_{\mathrm{st},j} = \frac{{\!\!\!\mathbf{\widehat{h}}^{\left( t \right)}_{j}}}{{{{\left\| {\mathbf{\widehat{h}}^{\left( t \right)}_{j}} \right\|}_2}}}.
\end{equation}

%After normalization, we send the normalized sample into the network training in turn to calculate the network output, and then obtain the output weight of hidden layer.
It is worth noting that the standardization operation is added before the activation function, which is given by
\begin{equation}\label{EQ10}
\mathbf{y} = {f_{\mathrm{std}}}\left( {\mathbf{W}{\mathbf{\widehat{h}}^{\left( t \right)}_{\mathrm{st},j}} + \mathbf{b}} \right),
\end{equation}
where ${f_{\mathrm{std}}}\left(  \cdot  \right)$ denotes the standardization operater with ${f_{\mathrm{std}}}\left( x \right) = {x \mathord{\left/
 {\vphantom {x {\left( {{{\mathds{E}}}\left( {\sqrt {{\mathds{D}}\left( x \right)} } \right)} \right)}}} \right.
 \kern-\nulldelimiterspace} {\left( {\mathds{E}\left( {\sqrt {\mathds{D}\left( x \right)} } \right)} \right)}}$. In this letter, we collectively refer to the normalization in (\ref{EQ9}) and the standardization operation in (\ref{EQ10}) as standardization. Then, the hidden layer output matrix ${\mathbf{o}^{\left( t \right)}_j}$ is expressed as
\begin{equation}\label{EQ11}
{\mathbf{o}^{\left( t \right)}_j} = \sigma \left( \mathbf{y} \right),
\end{equation}
where $\sigma \left(  \cdot \right)$ denotes the activation function such as sigmoid, hyperbolic tangent (tanh), rectified linear units (ReLU) \cite{c21}.
%In this letter, we select the sigmoid as the activation function since its output is mapped between $(0,1)$, the output range is limited.
By collecting ${\bm{\mathrm{O}} _j} = \left[ {\bm{\mathrm{o}} _j^{\left( 1 \right)}, \cdots ,\bm{\mathrm{o}} _j^{\left( {{N_d}} \right)}} \right]$, the hidden layer output weight ${{\bf\bm{\beta }}_j} $ is obtained by
 \begin{equation}\label{EQ12}
  {{\bf\bm{\beta }}_j} = {{\bf{H}}_j}{\bf{O}}_j^\dag.
  \end{equation}

%$2)$ \textit{Online Deployment}:
\subsubsection{Online Deployment}
 For the online deployment, the LS estimation given in Eq.~(\ref{EQ6}) is first employed to obtain the initial features of composite CFRs. Then, we use Eq.~(\ref{EQ7}) to separate the CFRs of direct link ${\mathbf{\widehat{h}}_{\mathrm{TR}}}$ and reflecting link ${\mathbf{\widehat{H}}_{\mathrm{TRR}}}$. According to the separated CFRs (e.g., ${\mathbf{\widehat{h}}_{\mathrm{TR}}}$ and ${\mathbf{\widehat{H}}_{\mathrm{TRR}}}$), we normalize the separated CFRs to form ${{ \mathbf{{{\overline{h}}}}}_{\mathrm{st},j}} $,
 i.e., ${{{\bf{\bar h}}}_{\mathrm{st},j}} = {{{{{\bf{\bar h}}}_j}} \mathord{\left/
 {\vphantom {{{{{\bf{\bar h}}}_j}} {\left( {{{\left\| {{{{\bf{\bar h}}}_j}} \right\|}_2}} \right)}}} \right.
 \kern-\nulldelimiterspace} {\left( {{{\left\| {{{{\bf{\bar h}}}_j}} \right\|}_2}} \right)}}$. And then the network output ${{ \mathbf{\widetilde{h}}}_j}$ is obtained by feeding ${{ \mathbf{{{\overline{h}}}}}_{\mathrm{st},j}} $ into the trained ELM-based network, which is given by
\begin{equation}\label{EQ13}
{\mathbf{\widetilde{h}}_j} = {\bm{\beta} _j} \,   .\sigma \left( {{f_{\mathrm{std}}}\left( {\mathbf{W}{{ \mathbf{{{\overline{h}}}}}_{\mathrm{st},j}}  + \mathbf{b}} \right)} \right).
\end{equation}

\section{Simulation Results and Analysis}

In this section, the normalized mean square error (NMSE) is applied as the metric to compare the performance between the proposed method and the recent novel methods, i.e., \cite{c25}, \cite{c24} and \cite{c1} with the considerations of insufficient CP and nonlinear distortion\footnote{We respectively employ the same architectures of \cite{c25} and \cite{c24} for simulating their NMSEs. And the source code is available on the authors link: https://github.com/meuseabe/deepChannelLearning4RIS}. The structure of this section is given as follows: The comparison of NMSE performance is given in Section IV-\textit{A}. Then, in Section IV-\textit{B}, the robustness of improvement with different parameters is discussed.
%At last, the generalization is analyzed in Section IV-\textit{C}.

The basic parameters involved are listed below. The training sequence is Zadoff-Chu sequence \cite{c1}, $L = 12$, $N = 64$, ${L_{\mathrm{CP}}} = 8$, and $M = 8$. The training and testing of the proposed method are carried out on a server with Intel Xeon(R) E5-2620 CPU 2.1GHz$\times$16, and the results are obtained by using MATLAB simulation on the server CPU due to the lack of a GPU solution. And the offline training takes about 13 seconds, whereas the online deployment of the proposed ELM network only needs 10 seconds. The frequency-selective Rician fading channels are considered.
%where the first tap is set as the deterministic line-of-sight (LOS) component and the remaining taps are non-LOS components following the Rayleigh distribution, with $\eta  = 0.5$ being the ratio of the total power of non-LOS components to that of LOS component \cite{c1}.
In this letter, we consider the effects of HPA, the nonlinear amplitude $A\left( x \right)$ and phase $\Phi \left( x \right)$ are respectively adopted from
\begin{equation}\label{EQ_nonlinear}
A\left( x \right) = \frac{{{\alpha _a}x}}{{1 + {\beta _a}{x^2}}}.{\rm{ }}\Phi \left( x \right) = \frac{{{\alpha _\phi }{x^2}}}{{1 + {\beta _\phi }{x^2}}},
\end{equation}
where ${\alpha _a} = 1.96$, ${\beta _a} = 0.99$, ${\alpha _\phi } = 2.53$, and ${\beta _\phi } = 2.82$ are considered in the simulations.

%\begin{table}[t]
%%\begin{table}0
%\renewcommand{\arraystretch}{1}
%\caption{Abbreviation}
%\label{tableQ_1} \centering
%%\begin{tabular}{l}
%%\hline
%%\\
%\begin{tabular}{c|c|c|c|c}
%\Xhline{1.0pt}
%\ & Ref\cite{c1}   & Prop & Ref\cite{c1} + Dis & Prop + Dis \\
%\Xhline{1.0pt}
%Reference method given in \cite{c1}   & $\surd$  & $\times$  & $\surd$  & $\times$    \\
%\hline
%Proposed eELM-CE method & $\times$         & $\surd$    & $\times$   & $\surd$ \\
%\hline
%Insufficient CP & $\surd$     & $\surd$    &$\surd$    & $\surd$ \\
%\hline
%Nonlinear distortion & $\times$     & $\times$    &$\surd$    & $\surd$ \\
%\hline
%
%\end{tabular}
%\end{table}

In addition, the intensity of nonlinear distortion is measured by error vector magnitude (EVM), which is defined as
\begin{equation}\label{SNR13}
{\mathrm{EVM} \left(\%\right) = \sqrt {\frac{{\sum\limits_{n \in N} {{{\left| {{{\widetilde x}_n} - {R_n}} \right|}^2}} }}{{\sum\limits_{n \in N} {{{\left| {{R_n}} \right|}^2}} }}}  },
\end{equation}
where ${\widetilde{x}_n}$ and ${R_n}$ are the distorted and undistorted outputs of HPA given the same input, respectively. That is, for the same input, ${\widetilde{x}_n}$ and ${R_n}$ are the outputs when HPA works in the saturated region and linear region, respectively. Except for the robustness analysis against EVMs, we set the basic value of EVM as ${55\%}$.

For simplicity, we use ``Ref\cite{c25}", ``Ref\cite{c24}", ``Ref\cite{c1}" and ``Prop" to denote the reference method given in \cite{c25}, \cite{c24}, \cite{c1} and the proposed eELM-CE method under nonlinear distortion with insufficient CP, respectively.

\subsection{NMSE Performance Analysis}
\begin{figure}[t]
\centering
\includegraphics[scale=0.5]{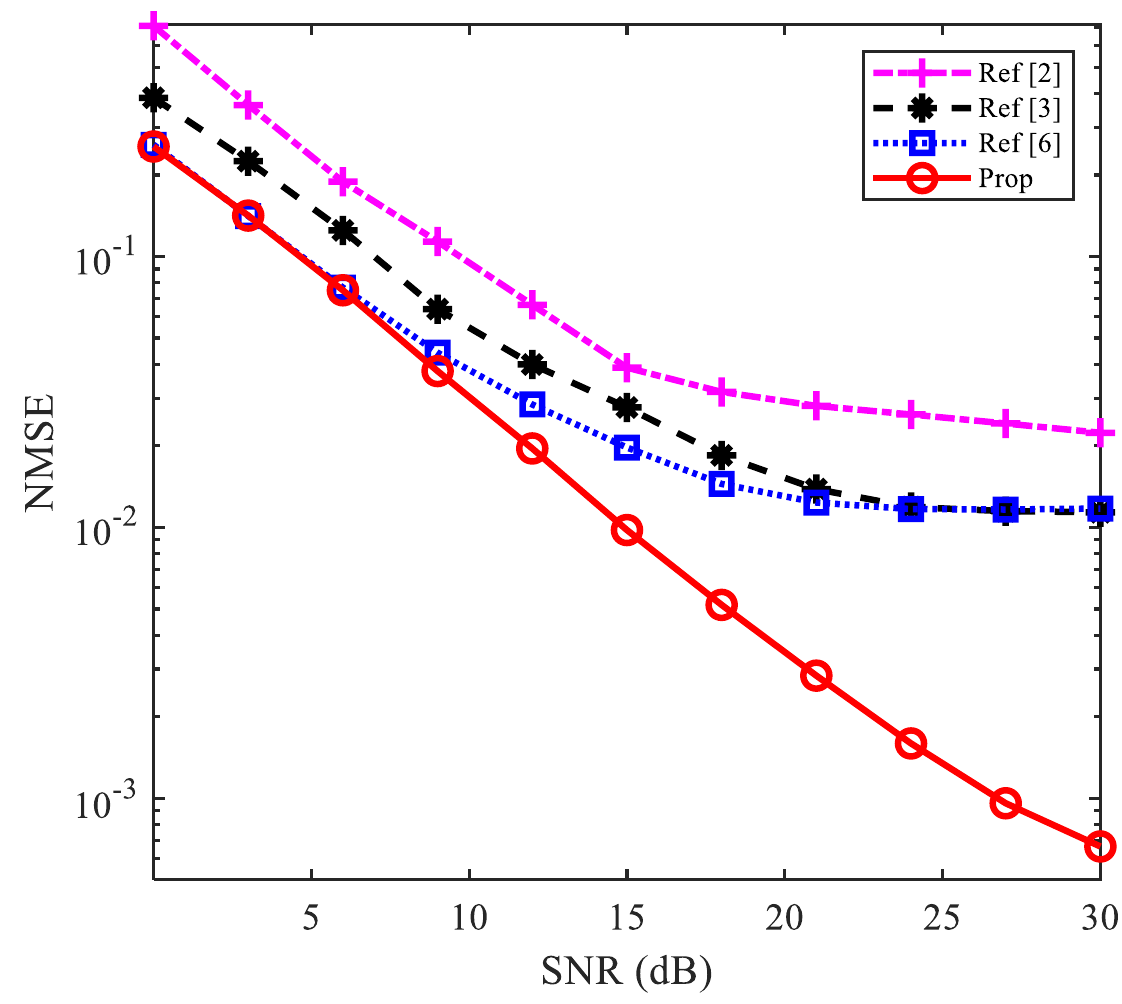}
\caption{ NMSE vs. SNR with insufficient CP. }
%\captionsetup{justification=centering}
\label{figeff_inCP}
\end{figure}

We add the nonlinear distortion and insufficient CP on the basis of \cite{c1}, which fully proves the effectiveness of the proposed method in terms of the NMSE curves in Fig.~\ref{figeff_inCP}. As shown in Fig.~\ref{figeff_inCP}, the NMSE performance of ``Ref\cite{c25}", ``Ref\cite{c24}" and ``Ref\cite{c1}" is  much higher than that of ``Prop". It is noticed that the proposed method has similar or better NMSE performance than ``Ref\cite{c25}'', ``Ref\cite{c24}'' and ``Ref\cite{c1}'' during all SNR range. This is mainly due to the considerations of insufficient CP and the nonlinear distortion as well as the application of enhanced ELM. Besides, different from the data-driven mode adopted by ``Ref\cite{c25}'' and ``Ref\cite{c24}'', the model-driven mode is employed by ``Prop" to exploit the advantages of both conventional algorithms and data-driven mode. In addition, since improving the ELM network is the principal task, we focus on the effectiveness of the proposed method, and the dedicated scenarios in ``Ref\cite{c25}'' and ``Ref\cite{c24}'' are not explored.

%since we focus on the effectiveness of the proposed method, and the dedicated scenarios in ``Ref\cite{c25}'' and ``Ref\cite{c24}'' are not explored.

%Since this letter is to improve the ELM network, we focus on the effectiveness of the proposed method,

%,  e.g., channel gain, DOA and AOA, are not the main consideration in our work, and thus cannot obtain the contributions from dedicated scenarios.
%Besides, we also observe that the performances of ``Ref\cite{c25}" and ``Ref\cite{c24}" saturate at high SNR (e.g., $\mathrm{SNR}>15dB$) because of the biased nature of the neural networks which do not provide unlimited accuracy.
%Therefore, the proposed eELM-CE method shows improvement of NMSE performances in the distorted scenarios.

\subsection{Robustness Analysis}\
Usually, the NMSE performance of channel estimation is influenced by the number of multi-path (i.e., $L$) and the degree of EVM. It is worth noting that, besides the change of the impact parameter (i.e., $L$ and EVM), other basic parameters remain the same as those given in Section IV.
%during the simulations.
\begin{figure*}[t]
\centering
\includegraphics[scale=0.5]{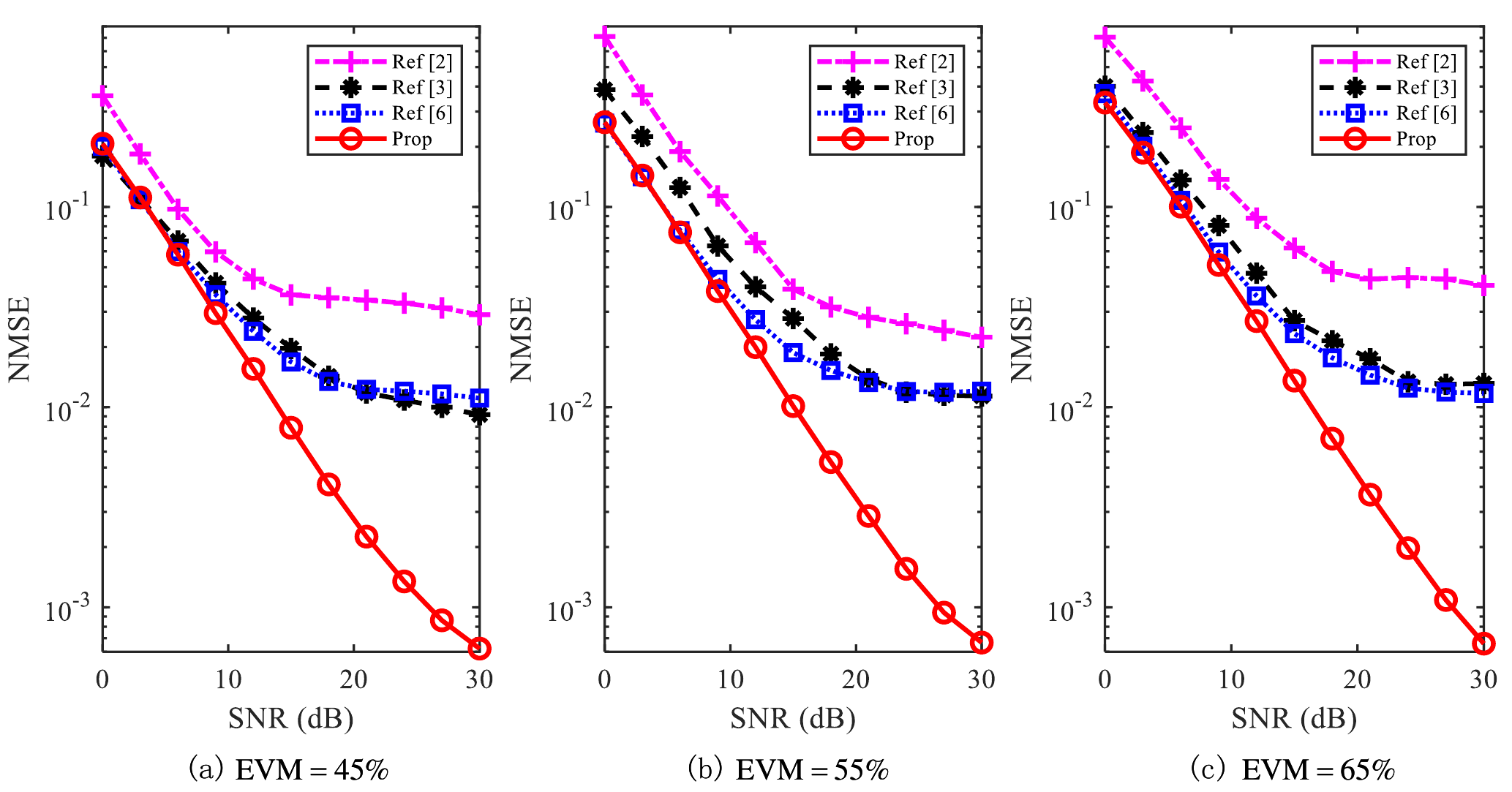}
\caption{ NMSE vs. SNR with different EVMs.}
%\captionsetup{justification=centering}
\label{evm}
\end{figure*}

\subsubsection{Robustness against EVM}
\text{\\ \\ \\}
EVM is usually used to measure the distortion intensity. To analyze the robustness of the proposed method against different distortion intensities, Fig.~\ref{evm} plots the curves of NMSE with different EVMs (i.e., ${\rm{EVM}} = 45\% $, ${\rm{EVM}} = 55\% $ and ${\rm{EVM}} = 65\% $). From Fig.~\ref{evm}, with the increase of EVM, the NMSEs for all curves increase due to the rise of distortion intensity. Besides, compared with ``Ref\cite{c25}", ``Ref\cite{c24}" and ``Ref\cite{c1}", the ``Prop" method achieves the smallest NMSE for each given EVM. As a result, against the impact of EVM, the ``Prop" possesses its robustness for improving the performance of NMSE.
\begin{figure*}[t]
\centering
\includegraphics[scale=0.5]{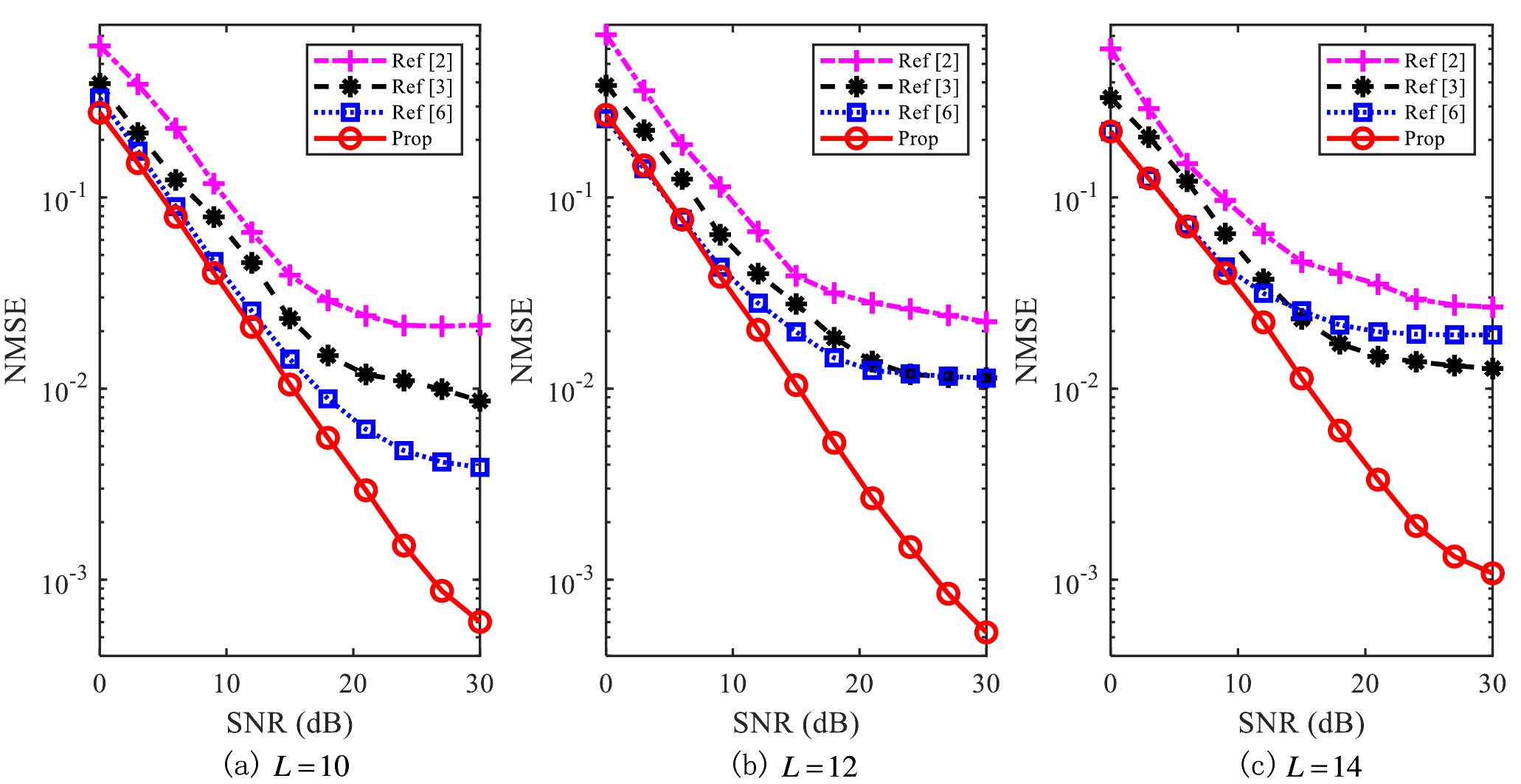}
\caption{ NMSE vs. SNR with different numbers of multi-path $L$.}
%\captionsetup{justification=centering}
\label{figpath}
\end{figure*}
\subsubsection{Robustness against $L$}
\text{\\ \\ \\}
In this paragraph, explicitly refer to Fig.~\ref{figpath}, it can be seen from the simulations that as the degree of insufficient CP increases, the performance of the NMSE deteriorates. For example, when ${L_{\mathrm{CP}}} = L - 2$, $\mathrm{SNR} = 15 \mathrm{dB}$, the reference methods \cite{c25}, \cite{c24} and \cite{c1} start to bend upward, but the performance is improved after using our proposed method, which reflects the robustness of our proposed method against $L$.

%
%\subsection{Generalization Analysis}
%In Fig.~\ref{genalization}, the trained network of $ L = 8 $ is employed to examine the tested network of $L=8$ and $L=6$. The performance of NMSE is degraded when the testing $L$ is not the training $L$ using eELM-CE. Besides, when $L\_{\rm{train}} = L\_{\rm{test}}$, the performance of the reference method \cite{c1} deteriorates because the degree of insufficient CP increases, but nevertheless the NMSE of using eELM-CE method is obviously lower than that of the reference method \cite{c1}.
% \begin{figure}[ht]
%\centering
%\includegraphics[scale=0.60]{genalization.pdf}
%\caption{ NMSE vs. SNR aganist $L$. }
%\captionsetup{justification=centering}
%\label{genalization}
%\end{figure}
\section{Conclusion}
In this letter, a channel estimation method, eELM-CE is proposed to improve the performance of the RIS-assisted OFDM system affected by the insufficient CP and imperfect hardware. Based on the model-driven mode, LS estimation is first employed as a feature extractor to extract the initial features. Then, with the extracted initial features, an ELM network is developed to refine the channel estimation. Noteworthily, the proposed ELM networks are enhanced by using hidden-layer standardization to better capture the features of distortion and thus enhance the performance. Extensive simulation results show the effectiveness of the proposed method in relieving the influence of insufficient CP and suppressing the nonlinear distortion. The robustness of eELM-CE is also validated by its stable performance against parameter variations. In this letter, the difficulty of obtaining desired labels is simplified by generating them according to the existing channel model. In our future works, we will consider the desired labels in real channel scenarios to promote the application of ELM-based channel estimation in practical systems (e.g., IoT, WLAN, etc).


%\end{IEEEbiography}


%\newpage



%\vfill


\end{document}